# Thermal and Magnetic Properties of Nanostructured Ferrimagnetic Composites with Graphene – Graphite Fillers

S. Ramirez[1,2], K. Chan[2,3], R. Hernandez[1], E. Recinos[2,3], E. Hernandez[1], R. Salgado[1], A.G. Khitun[2], J. E. Garay[2,3] and A. A. Balandin[1,2,1]

[1]Nano-Device Laboratory (NDL), Department of Electrical and Computer Engineering and Phonon Optimized Engineered Materials (POEM) Center, University of California – Riverside, Riverside, California 92521 USA

[2]Spins and Heat in Nanoscale Electronic Systems (SHINES) Center, University of California-Riverside, Riverside, California 92521 USA

[3]Advanced Materials Processing and Synthesis (AMPS) Laboratory, Department of Mechanical Engineering, University of California – Riverside, Riverside, California 92521 USA

---

[1] Corresponding author (A.A.B.): balandin@ece.ucr.edu

## Abstract

We report the results of an experimental study of thermal and magnetic properties of nanostructured ferrimagnetic iron oxide composites with graphene and graphite fillers synthesized via the current activated pressure assisted densification. The thermal conductivity was measured using the laser-flash and transient plane source techniques. It was demonstrated that addition of 5 wt. % of equal mixture of graphene and graphite flakes to the composite results in a factor of ×2.6 enhancement of the thermal conductivity without significant degradation of the saturation magnetization. The microscopy and spectroscopic characterization reveal that $sp^2$ carbon fillers preserve their crystal structure and morphology during the composite processing. The strong increase in the thermal conductivity was attributed to the excellent phonon heat conduction properties of graphene and graphite. The results are important for energy and electronic applications of the nanostructured permanent magnets.

## I. INTRODUCTION

Magnetic nanocomposites are important for a wide range of applications, and they are playing an increasingly important role for uses in medicine [1-3], catalysis [4] and other applications [5]. The high concentration of interfaces in nanocomposites allows one to take advantage of different magnetic phases for better magnetic coupling, e.g., ferrimagnetic – antiferromagnetic, ferromagnetic – nonmagnetic. However, the operation temperature of nanostructured magnets can be severely limited owing to poor heat conduction properties of nanostructured materials, resulting in temperature rise and degradation of magnetic properties, which are highly temperature sensitive. Permanent magnets, used in a variety of applications, ranging from power sources to switches and actuators [6-7], can be exposed to temperatures higher than or approaching the Curie temperature. For this reason, thermal management of nanocomposite magnets is becoming an important consideration. The magnetization of magnetic materials decreases as the Curie point is reached. A conventional approach of avoiding it is increasing the Curie point via materials engineering and utilization of nanostructured materials [8-9]. However, another strategy – increasing the thermal conductivity, $K$, and thermal diffusivity, $\alpha$, of permanent magnets – has not been well studied. Indeed, higher $K$ and $\alpha$ would translate into lower temperature rise allowing for functioning of practical systems based on magnets with relatively low Curie temperatures. This approach may be particularly relevant to nanostructured magnets that have very low thermal conductivity owing to the heat carriers, acoustic phonons or electrons, scattering on grain boundaries and defects. An obstacle for improving heat conduction properties of nanostructured magnets is a possibility of degrading the magnetic characteristics. Thus, one should attempt to increase the thermal conductivity without substantial degradation of magnetization.

In this work, we show that the use of graphene and graphite flakes as fillers in properly synthesized nanostructured materials allows one to significantly increase the heat conduction properties of nanostructured magnets without serious degradation of their magnetization. As a model system for the proof-of-concept demonstration of our approach we used the densified nanostructured iron oxide. The bulk nanostructured samples for this study were synthesized via the current activated pressure assisted densification (CAPAD) process also known as the spark plasma sintering [10-12]. Iron oxide was chosen as a model system due to its natural abundance, ease of processing, and extensively studied properties. It is likely that the demonstrated approach of utilizing thermal fillers will work with other nanostructured magnetic materials. As an additional reference sample, we used a free-sintered sample (without densification) to illustrate the importance of proper steps in composite preparation for preserving magnetic properties.

As the thermal filler material we selected a mixture of graphene and graphite flakes owing to their excellent thermal properties [13-14]. High quality graphite has the in-pane thermal conductivity $K$=2000 W/m·K at room temperature (RT). The intrinsic thermal conductivity of graphene can be even higher than that for bulk graphite basal planes [13]. Graphene and few-layer graphene (FGL) flakes were shown to work well as thermal fillers in various polymer composites used as thermal interface materials (TIMs) [15-18]. In many cases, FLG fillers perform better than graphene or graphite particles because they preserve the mechanical flexibility of graphene and do not suffer from degradation of the intrinsic heat conduction properties while in close contact with the matrix materials. However, there have been no reports of graphene, FLG or graphite thermal fillers in magnetic materials or other related mechanically hard solids. We focused our study on finding an optimum mixture of graphene and graphite fillers for nanostructured permanent magnets. Since graphene has a thickness of one atomic plane its high temperature processing with iron oxide can result in chemical reaction and alloying with a corresponding degradation of thermal properties. Graphite flake with lower surface to volume ratio are expected to be more robust for preserving their morphology and intrinsic heat conduction properties. At the same time certain fraction of graphene can be beneficial for forming heat conduction network owing to graphene flexibility and excellent coupling to matrix materials. The latter was observed in the epoxy-based TIMs where mixtures of graphene and FLG or graphene and graphite performed the best [15-17]. The cost considerations also favor the use of graphite – FLG mixtures rather than pure graphene or FLG solutions obtained by the liquid phase exfoliation or graphene oxide reduction.

## II. MATERIAL SYNTHESIS AND CHARACTERIZATION

In order to synthesize the samples, we used commercial 20-nm grain size $\gamma$-Fe$_2$O$_3$ powder (Inframat Advanced Materials) and graphite (Asbury Carbons). The materials were dispersed into ethanol and then ultra-sonicated for one hour. The ethanol was evaporated for 12 hours at 80 ºC and collected. In the CAPAD process the powder was rapidly densified at 600 ºC with an applied load of 150 MPa [10-11]. The samples with the mixture of graphite and graphene were prepared by the same approach using commercial graphene – FLG solution (Graphene Supermarket). For comparison, we prepared several samples using traditional sintering without densification. The morphology and crystal structure of the samples were studied using scanning electron microscopy (SEM), Raman spectroscopy (Renishaw) and X-ray diffraction (XRD). Figure 1 (a-b) shows SEM image of fracture surface of a representative ferrimagnetic iron oxide composites with graphite and graphene fillers at two distinct magnifications. Analysis of the microscopy data indicate that the samples consist of iron oxide grains and separate graphite flakes (50 – 800 µm diameters, 1 – 150 µm thick) together with graphene flakes of various lateral dimensions. The microcopy data do not show obvious signs of reaction between the iron oxide grains and graphite flakes, *i.e.* the CAPAD process preserved morphology.

[Figure 1 (a-b): SEM]

In Figure 2 we present Raman spectrum of a composite sample. The G peak and 2D band, characteristic for both graphite and graphene, are clearly observed approximately at ~1580 cm$^{-1}$ and ~2700 cm$^{-1}$, respectively. They confirm the graphitic sp$^2$ crystal structure of the carbon fillers after thermal processing. The Raman peaks observed between ~200 cm$^{-1}$ and ~700 cm$^{-1}$ are in the same range as the phonon bands reported in literature for magnetite ($Fe_3O_4$) [19-22]. However, one should keep in mind that there is substantial variation in the Raman peak positions and their assignment for magnetite [19]. The discrepancy is usually attributed to the defects, sample surface conditions, inclusions of hematite (α-$Fe_2O_3$) and local heating by the excitation laser. In our case, the iron oxide samples are composites of magnetite and hematite. The nanostructured nature of the iron oxide and its intermixture with carbon fillers can also affect the spectral position and broadening of the phonon peaks. The iron oxide magnetite crystallizes in the cubic space group *Fd*3m ($O_h^7$). The theoretical analysis based on the factor-group approach predicts five Raman-active bands $A_{1g}$, $E_g$ and three $T_{2g}$. Previous studies of the iron oxide magnetite crystals revealed theoretically predicted phonon bands at 193 cm$^{-1}$, 306 cm$^{-1}$, 538 cm$^{-1}$ and 668 cm$^{-1}$ [19]. In the spectrum of our samples, we can also identify $A_{1g}$, $E_g$ and $T_{2g}$ bands and relatively close coincidence of some of the peaks.

[Figure 2: Raman]

The combined SEM and Raman inspection of the samples demonstrated clearly the separation of the iron oxide and carbon phases. No other phases such as carbon steel or carbide formed during the CAPAD processing. To further confirm that we prevented reaction between the graphite fillers and iron oxide via the rapid densification capabilities of CAPAD we conducted X-ray diffraction (XRD) study of the pure iron oxide and iron oxide – carbon composites (Figure 3). As expected the iron oxide samples shows only iron oxide peaks, cubic iron oxide ($\gamma$-$Fe_2O_3$ or $Fe_3O_4$) and hexagonal ($\alpha$-$Fe_2O_3$). From previous studies [23] we know that densification via CAPAD leads to mixed iron oxide phases (cubic + hexagonal) due to the reducing processing environment. Both composite samples (iron oxide + 5% graphite and iron oxide + 2.5% graphite + 2.5% graphene) show cubic and hexagonal iron oxide peaks. As expected these samples also show clear carbon peaks.

[Figure 3: XRD]

Cubic iron oxides ($\gamma$-$Fe_2O_3$ or $Fe_3O_4$) are ferrimagnetic and find uses as soft permanent magnets with a high saturation magnetization, typically between 60-80 emu/g, and low coercivity, while hexagonal $\alpha$-$Fe_2O_3$ is an antiferromagnetic material [24]. In an ideal magnetic composite the total saturation magnetization should obey a rule of mixtures by decreasing proportionally with the amount of non-magnetic material added [11, 23-25]. For this reason, we limited the amount of carbon thermal fillers to below 10 wt. %. The magnetic hysteresis of our samples was measured by the vibrating sample magnetometry (VSM) system (Lakeshore 7400 Series). In this measurement, the VSM system rapidly vibrates a sample between two pickup coils while applying a magnetic field. The created alternating magnetic field induces an electric bias, which defines the magnetic moment of the sample. The samples were initially magnetized by sweeping the applied field from zero to 17 kOe. The full hysteresis loop was obtained from -17 kOe to 17 kOe.

Figure 4 shows the magnetic properties of a pure iron oxide reference sample and the CAPAD processed iron oxide composite with an addition of 5 wt. % of graphite (red line), and iron oxide composites with an addition of 2.5 wt. % of graphite and 2.5 wt. % of graphene (green line). The data for the free-sintered, not CAPAD densified, iron oxide sample with graphite is also shown for comparison (blue line). The most important observation is that the ferrimagnetic properties of CAPAD processed iron oxides retained after addition of graphite or graphene thermal fillers. The small decrease in the saturation magnetization, $M_s$ of the iron oxide + 5% graphite compared to the pure iron oxide sample is consistent with the dilution effects expected in a ferrimagnetic – diamagnetic composite system. A somewhat stronger decrease of the magnetization in the iron oxide with 2.5% graphite and 2.5% graphene loading is attributed to the increased hexagonal (antiferromagnetic) content this sample, which further dilutes the high $M_s$ of the cubic ferrimagnetic phase. The data in Figure 4 suggest that addition of graphite and graphene fillers to the nanostructured ferrimagnetic materials via CAPAD processing did not severely degrade the magnetization. The CAPAD process preserved the separate phases of graphite and magnetic materials. Lack of reaction products is important given that iron and carbon readily form carbides. It is also interesting to note that the samples obtained via conventional free-sintering methods have greatly reduced saturation magnetization, and demonstrate primarily antiferromagnetic behavior, characteristic of hexagonal α-$Fe_2O_3$. They also have significantly lower density so that the resultant samples do not lend themselves well to applications.

[Figure 4: Magnetic]

### III. THERMAL MEASUREMENTS

The thermal conductivity of the samples was measured by two different techniques. The cross-plane thermal conductivity (perpendicular to the sample plane) was determined using the laser flash technique (LFT) (Netzsche) in the temperature range from 20ºC to 100ºC. This method involves a measurement of the thermal diffusivity, $\alpha$, and a separate measurement of the specific heat, $C_p$ [26-27]. The thermal conductivity is calculated as $K=\alpha\rho C_p$, where $\rho$ is the mass density of the sample. The specific heat is determined using calorimetry measurements (DSC Polyma).

The in-plane thermal conductivity was measured by the transient plane source (TPS) technique (Hot Disk) [28-29]. The in-plane thermal measurements were conducted at room temperature (RT). The cross-plane thermal conductivity for a range of iron oxide samples with different fraction of graphite and graphene shown in Figure 5. The thermal conductivity of reference pure nanostructured iron oxide was measured to be 2.7 W/mK at RT. The important observation from Figure 5 is that the thermal conductivity of magnetic samples with carbon thermal fillers noticeably increases. The $K$ enhancement is the highest in the sample that contain a 2.5 wt. % graphene and 2.5 wt. % graphite mixture of thermal fillers. The RT thermal conductivity goes up from $K$=2.7 W/m·K to 6.0 W/m·K. The enhancement at elevated temperatures is even larger. One can also see that the increase of the thermal conductivity is a non-monotonic function of the graphite loading. The sample with 5 wt. % loading of graphite demonstrate larger $K$ than that with 10 wt. % graphite loading. In addition, the samples with higher than 5 wt. % of graphite had stronger degraded magnetization.

[Figure 5: Thermal]

Anisotropy in heat conduction in magnetic samples is expected due to the high aspect ratio of the graphite and graphene fillers, which may have predominant orientation owing to the specifics of the sample preparation. The TPS technique, used for the in-plane thermal conductivity measurement, utilizes two thin sensors that are placed between two halves of a sample under investigation. In this measurement, the sensor acts as both a heater and a thermometer [28-29]. It measures the temperature rise of the sample as a function of time. In the transient measurement both thermal diffusivity and thermal conductivity can be determined. The results of the measurements are summarized in Table I. In addition to $K$ values, we also show the thermal conductivity enhancement factor $TCE=(K-K_m)/K_m$, where $K_m$ is the thermal conductivity of iron oxide without thermal fillers. For comparison, the cross-plane thermal conductivity at RT is also listed. The nanostructured pure iron oxide has the in-plane thermal conductivity of ~2.6 W/m·K, which is consistent with the cross-plane value of 2.7 W/mK. This indicates isotropic heat transport in this type of magnetic materials without thermal fillers. The in-plane thermal conductivity increases to ~5.4 W/m·K (TCE of 110%) for the sample containing 5% graphite and increases to ~9.3 W/m ·K (TCE of 260%) for the sample containing graphite and graphene. The higher in-plane values of thermal conductivity were attributed to predominantly in-plane orientation of the graphite and graphene fillers after powder loading and sonication. The strong enhancement of the thermal conductivity (by a factor of ×2.6) with relatively small decrease in magnetization indicates a new promising way of thermal management of permanent magnets.

[Table I: Thermal]

## IV. DISCUSSION

We verified that the lattice vibrations (e.g. acoustic phonons) are the main heat carriers in the nanostructured ferrimagnetic iron oxide composites with graphitic fillers by determining their electrical resistivity. The measurements were performed using the four-probe Van der Pauw technique. The electrical resistivity, $\rho_e$, was in the range from 3.8 Ω–m to 5.8 Ω–m and did not change significantly with addition of carbon fillers with the loading fraction below 5 wt. %. The electronic contribution, $K_e$, to the total thermal conductivity, $K$, was estimated from the Wiedemann-Franz law: $K_e/\sigma=(\pi^2/3)(k_B/e)^2T$, where $\sigma=1/\rho_e$ is the electric conductivity, $k_B$ is the Boltzmann constant and $e$ is the charge of an electron. Our data indicates that $K_e$ was less than 1% of the total thermal conductivity. These measurements also suggest that graphene and graphite fillers do not form a percolation network at the studied sp$^2$ carbon loading fractions. The acoustic phonons conduct heat partially via the graphite and graphene fillers and partially via iron oxide grains. One should note here that the concept of the phonon as extended state (plane wave) in the context of highly disordered materials has its inherent limitations. Other theoretical models such as thermal excitation hopping transport may be more appropriate [30-31].

We now compare the thermal conductivity data for nanostructured iron oxide with that of bulk iron oxide crystals reported in literature. It should be noted that there is not a significant amount of previous work on thermal conductivity of iron oxide. A previous study found that Fe$_3$O$_4$ polycrystals prepared by the hot isostatic pressing (HIP) method had the thermal conductivity approximated by the empirical equation [32]: $K=1/(1.844\times10^{-4}T)$ for the temperature in the range 298 K $< T <$ 912 K. At RT, this formula gives $K \approx 20$ W/m·K. The thermal conductivity of bulk crystals, according to this equation, decreases as $K \sim 1/T$. This is consistent with the common trend in electrical insulator and semiconductor crystals, where the thermal conductivity is limited by the crystal lattice anharmonicity, i.e. phonon Umklapp scattering. An earlier experimental study on single crystals reported that the thermal conductivity of magnetite crystals follows the equation [33]: $K=0.0423 - 1.37 \times 10^{-5}T$ for the temperature range from 340 K to 675 K (in this equation, $K$ is in W/cm·K). The RT value given by this equation is about 3.8 W/m·K. The same study found that under pressure the thermal conductivity of magnetite increases. The cause of this significant difference between HIP samples [32] and the single crystals [33] is not clear. One would expect the $K$ of the single crystal samples to be higher since the HIP samples contain grain boundaries, which are well known to scatter phonons and electrons.

The thermal conductivity values measured in this work for nanostructured iron oxides samples are similar to those reported in Ref. [33]. The temperature dependence of thermal conductivity established for our samples is different from that in previous measurements [32]. The thermal conductivity in our samples increases with temperature, which is characteristic for disordered or nanostructured materials [30-31]. It is also clear from our experiments that the thermal conductivity in the nanostructured iron oxide composites with graphite and graphene fillers cannot be explain by the effective medium approximation (EMA) assuming the bulk values of the thermal conductivity of the constituents and their mass fractions. The thermal conductivity of the composites does not monotonically increase with the graphite loading fraction and the $K$ values for iron oxide and graphite – graphene fillers are size limited. The optimum concentration of $sp^2$ carbon (graphite and graphene) fillers has to be determined for each specific magnetic material. It also depends on the composite processing parameters.

## V. CONCLUSIONS

We reported on the thermal and magnetic properties of nanostructured ferrimagnetic iron oxide composites with graphene and graphite fillers synthesized via CAPAD. It was demonstrated that addition of 5 wt. % of equal mixture of graphene and graphite fillers to the composite results in a factor of ×2.6 enhancement of the thermal conductivity without significant degradation of the saturation magnetization. The microscopy and spectroscopic characterization reveal that carbon fillers preserve their crystal structure and morphology during the CAPAD composite processing. The results are important for energy and electronic applications of nanostructured magnetic composites. The nanostructured magnetic composites that have particularly low thermal conductivity owing to the heat carrier scattering on grain boundaries. Improved thermal management of such magnets may substantially increase their application domain.

*Acknowledgements*

The work at UC Riverside was supported as part of the Spins and Heat in Nanoscale Electronic Systems (SHINES), an Energy Frontier Research Center funded by the U.S. Department of Energy, Office of Science, Basic Energy Sciences (BES) under Award # SC0012670. The authors thank Prof. J. Shi and Prof. R. Lake for useful discussions.

**FIGURE CAPTIONS**

**Figure 1 (a-b):** Scanning electron microscopy micrograph of nanostructured iron oxide – graphite – graphene composite at two magnifications. Note the separation of the $sp^2$ carbon fillers and iron oxide grains indicating that no chemical reaction happened during the composite processing. The samples were prepared using CAPAD method.

**Figure 2:** Raman spectrum of the nanostructured iron oxide – graphite – graphene composite with clearly visible phonon bands of $sp^2$ carbon at 1580 $cm^{-1}$ and 2700 $cm^{-1}$ and those associated with iron oxide in the range from ~200 $cm^{-1}$ to ~700 $cm^{-1}$. The peaks confirm the separation of the iron oxide and carbon phases. The samples were prepared using CAPAD method.

**Figure 3:** XRD spectra of the nanostructured iron oxide – graphite – graphene composite. The peaks indicate separation of phases. The samples were prepared using CAPAD method.

**Figure 4:** Magnetic hysteresis curves for pure iron oxide , magnetite composite with graphite fillers prepared using CAPAD, magnetite composite with graphite and graphene fillers prepared using CAPAD, and iron oxide with graphite fillers prepared via traditional free-sintering. Note the strongly degraded saturation magnetization in the free-sintered sample and retained saturation in the CAPAD prepared sample.

**Figure 5:** Cross-plane thermal conductivity of nanostructured iron oxide – graphite composite samples containing 0%, 5%, 7%, and 10% of graphite fillers. The data is also shown for the nanostructured iron oxide – graphite – graphene composite with 2.5% of graphite and 2.5% of graphene. Note that the increase of the thermal conductivity is the highest in the composite with a mixture of graphite and graphene thermal fillers.

**Table I:** Thermal Conductivity of Nanostructured $Fe_3O_4$ with Graphitic Thermal Fillers

| Sample | Loading Fraction (%) | K (W/mK) | TCE (%) |
|---|---|---|---|
| Iron oxide | 0 | 2.7 (cross-plane) <br> 2.6 (in-plane) | 0 |
| Iron oxide + graphite | 5 | 3.6 (cross-plane) <br> 5.4 (in-plane) | 29 (cross-plane) <br> 110 (in-plane) |
| Iron oxide + graphite + graphene | 2.5 (graphite) + 2.5 (graphene) | 6.0 (cross-plane) <br> 9.3 (in-plane) | 120 (cross-plane) <br> 260 (in-plane) |

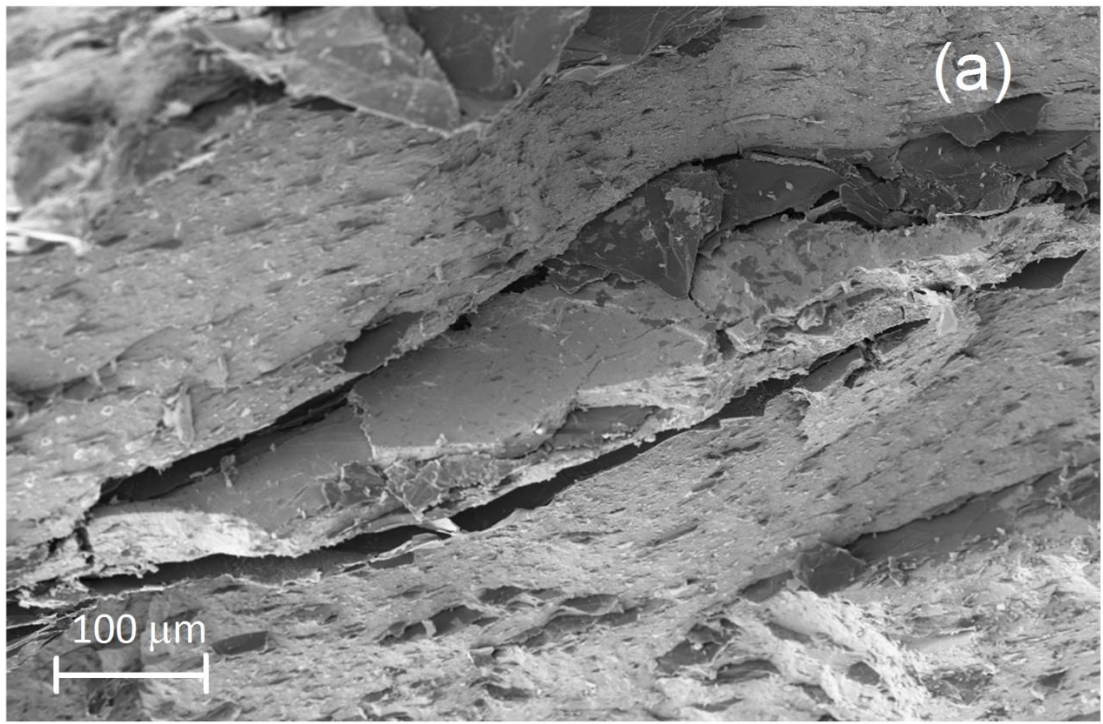
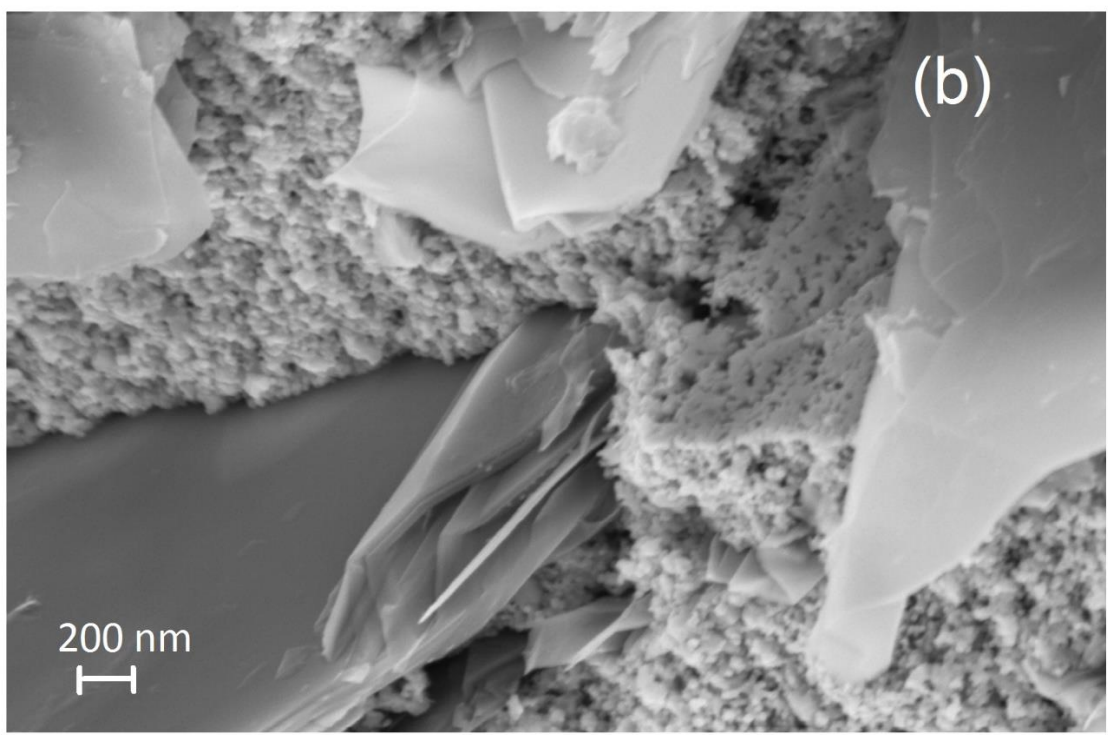

Figure 1

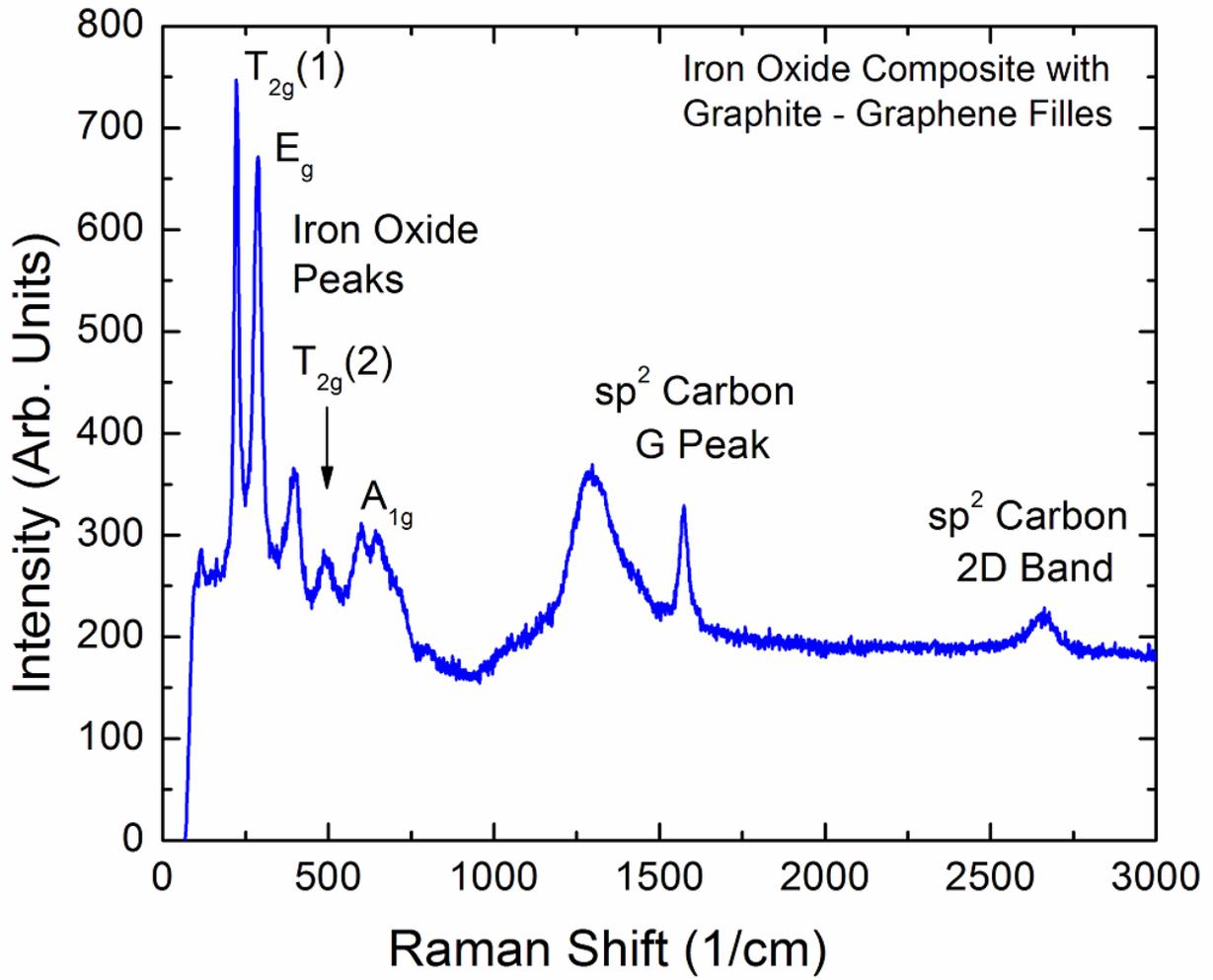

Figure 2

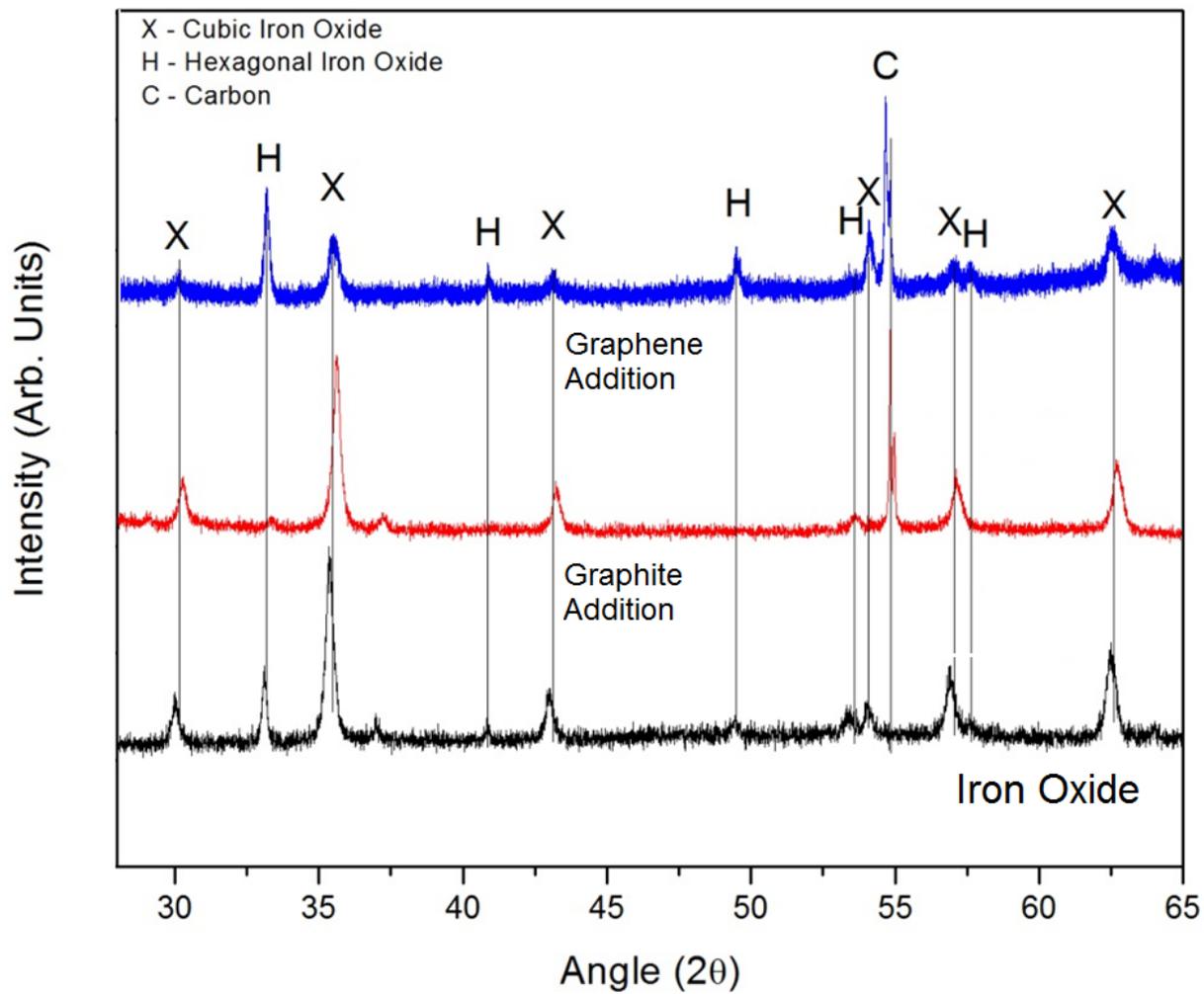

Figure 3

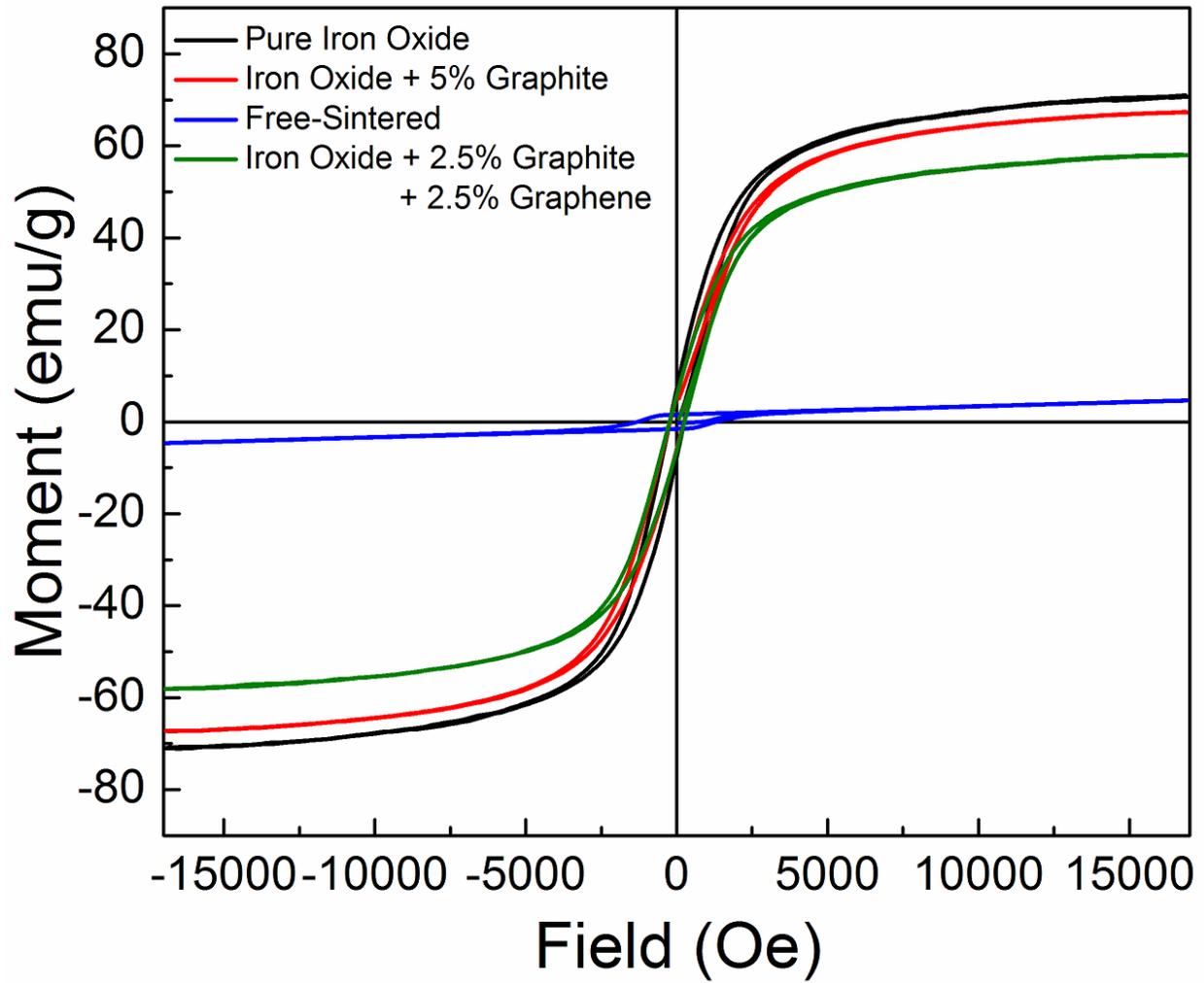

Figure 4

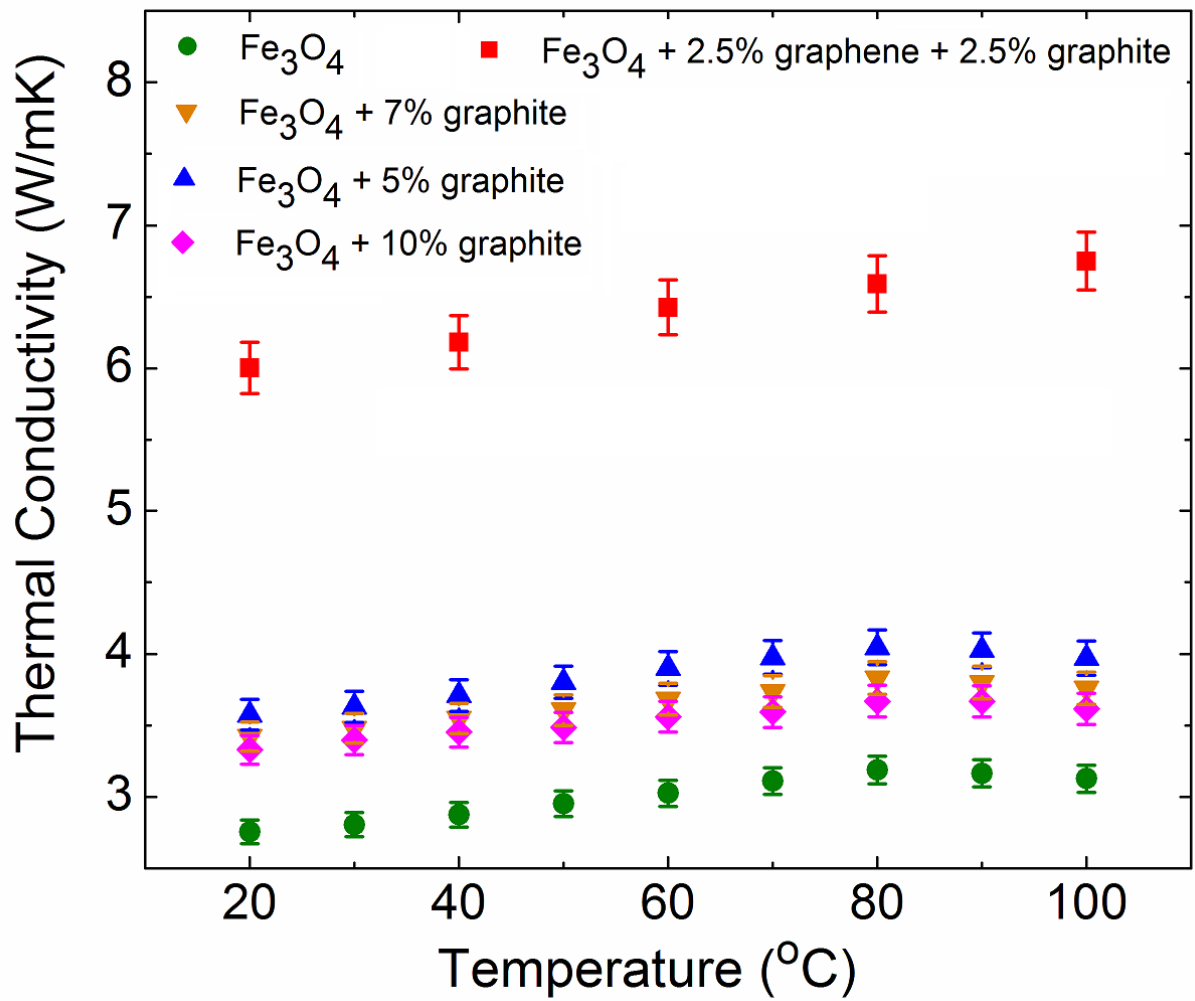

Figure 5